\pdfoutput=1
\documentclass[aps,prl,twocolumn,groupedaddress, showpacs]{revtex4}
\usepackage{graphicx}

\begin{document}


\title{Observation of vortex nucleation in a rotating two-dimensional lattice of Bose-Einstein condensates.}


\author{R.\ A.\ Williams}
\author{S.\ Al-Assam}
\author{C.\ J.\ Foot}
\affiliation{Clarendon Laboratory, Department of Physics, University of Oxford, Parks Road, Oxford, OX1 3PU, United 
Kingdom}

\date{\today}

\begin{abstract}
We report the observation of vortex nucleation in a rotating optical lattice.  A $^{87}$Rb Bose-Einstein condensate was 
loaded into a static two-dimensional lattice and the rotation frequency of the lattice was then increased from zero.  
We studied how vortex nucleation depended on optical lattice depth and rotation frequency.  For deep lattices above the 
chemical potential of the condensate we observed a linear dependence of the number of vortices created with the 
rotation frequency, even below the thermodynamic critical frequency required for vortex nucleation.  At these lattice 
depths the system formed an array of Josephson-coupled condensates.  The effective magnetic field produced by rotation 
introduced characteristic relative phases between neighbouring condensates, such that vortices were observed upon 
ramping down the lattice depth and recombining the condensates. 
\end{abstract}


\maketitle

Ultracold quantum gases in rotating optical lattices find themselves at the intersection of two fields which have 
generated an impressive body of experimental and theoretical work.  The versatile and clean potentials offered by 
optical lattices have proved to be an incredibly adept system for exploring a wide range of fundamental problems in 
condensed matter physics such as the Mott-insulator transition \cite{greiner02}, Anderson localization \cite{billy2008, 
Roati08} and Tonks-Girardeau gases \cite{Kinoshita04}.  Similarly rotating quantum gases have provided an array of 
striking results, from the nucleation of vortex lattices in Bose gases \cite{aboshaeer01, schweikhard04} to rotating 
Fermi gases at the BCS-BEC crossover \cite{Zwierlein05}.  The close analogy between the physics of rapidly rotating 
neutral atoms and electrons under a magnetic field has led to considerable interest in the possibility of achieving 
strongly-correlated quantum Hall states in a rapidly rotating atomic gas \cite{wilkin2000, cooper01}.

Rotating optical lattices for cold atoms have generated a large amount of interest in their own right, which can 
broadly be divided into three main areas: (i) Weak rotating lattices have been used to pin vortices \cite{tung2006} and 
theoretical work has shown that rich vortex lattices structures, including lattices of doubly quantized vortices, are 
predicted to emerge \cite{kasamatsu06, bigelow05}; (ii)  Stronger optical lattice potentials with a large number of 
atoms per site (100-1000) realize the physics of Josephson junction arrays (JJA) under magnetic fields 
\cite{kasamatsu09,polini05}; (iii)  Arguably the most interesting regime is for a dilute gas in the tight binding 
regime in a rotating lattice where fractional quantum Hall physics is predicted to occur \cite{palmer:180407, 
sorenson05, bhat:043601}.  In all three regimes rich structures emerge that depend on the relative density of vortices 
and lattice sites.

We report on the first experiments with a rotating optical lattice in the `deep' lattice regime where lattices depths 
were reached such that a 2D array of weakly linked condensates was created, forming a Josephson junction array.  Under 
rotation this system realizes the uniformly frustrated JJA \cite{cooper08, kasamatsu09}.  We have investigated the 
nucleation of vortices in a rotating lattice, starting from a non-rotating condensate loaded into a static lattice, and 
then increasing the rotation rate of the lattice.  There has been one previous experiment employing a rotating optical 
lattice by Tung {\it{et al}}.\ \cite{tung2006} at JILA, where it was reported that heating due to mechanical 
instabilities and aberrations limited the rotating lattice to depths less than 30\% of the condensate's chemical 
potential.  We were not limited to this weak lattice regime and our system is markedly different to the JILA experiment 
where a vortex lattice was created by an elegant evaporative spin-up technique \cite{schweikhard04} before a weak 
co-rotating optical lattice was imposed.

The rotating lattice used in the experiments described in this paper was generated by dual-axis acousto-optic 
deflectors and a novel optical system, described in detail in \cite{Williams:08}.  The two-dimensional optical lattice 
was formed in the focal plane of a custom-made multi-element lens with a numerical aperture of 0.27.  Generating the 
rotation of the lattice acousto-optically gave intrinsically smooth rotation and allowed the rotation rate to be varied 
during an experimental run.

The Coriolis force on neutral atoms in the rotating frame has the same form as the Lorentz force experienced by charged 
particles in a magnetic field.  Rotation at angular frequency $\Omega$ gives rise to the effective magnetic vector 
potential $\mathbf{A = \Omega\times r}$, and the Peierls phase gained by an atom tunnelling from site $i$ to site $j$ 
is  $A_{ij}~=~(m\slash\hbar)\int_{\mathbf{r}_i}^{\mathbf{r}_j} \mathbf{A(r')}\cdot \mathrm{d}\mathbf{r'}$, where $m$ is 
the mass of the atom.  The sum of the link phases around a lattice plaquette of side $d$ obeys the property 
$\sum_{\mathrm{plaquette}} A_{ij} =  2\pi f$, where $ f = 2m\Omega d^2\slash h$ is the mean number of vortices per 
lattice plaquette.

The experiments started with a $^{87}$Rb condensate containing $2\times 10^5$ atoms with no visible thermal component, 
held in a trap formed by an axial symmetric Ioffe-Pritchard trap and a single red-detuned light sheet in the radial 
plane increasing the axial trapping frequency to give $\{\omega_r,\omega_z\} = 2\pi\{20.1, 53.0\}$~Hz.  The residual 
radial anisotropy of the combined magnetic and optical trap was $\omega_y\slash\omega_x = 1.008 \pm 0.003$.   The 
dipole sheet trap was creating using broadband light with a spectral width $\Delta\lambda \sim 3~\mathrm{nm}$ centred 
on 865~nm in order that the light had a short coherence length.  This ensured there were no interference effects, e.g.\ 
from multiple reflections at the vacuum cell wall, adding noise to the trapping potential.  With narrow-band light it 
was observed such perturbations acted to spin down the cloud and caused irregularities in the filling of the lattice.

The 2D optical lattice was formed in the radial plane of the trapped condensate by four circularly polarized beams at 
$\lambda = 830~\mathrm{nm}$ intersecting in the focal plane of a high N.A.\ objective lens, achieving a lattice 
constant of $2~\mu\mathrm{m}$.  One pair of beams was detuned by 10~MHz to the other pair ensuring interference between 
the orthogonal optical lattices did not affect the atoms.  The optical lattice was initially static as it was ramped to 
its final depth $V_0$ in the range $100~\mathrm{Hz} \leq V_0 \leq 4000~\mathrm{Hz}$ \cite{noteV0}.  The radial 
Thomas-Fermi radius of the condensate was $16~\mu\mathrm{m}$, resulting in the filling of $\sim 16$ lattice sites 
across the diameter of the condensate and around 200 sites in all.  The lattice enhanced interaction energy at the 
central sites rose above the chemical potential of the unperturbed condensate ($\mu \approx 500~\mathrm{Hz}$) meaning 
an array of individual condensates was not formed until $V_0 \sim 1000~\mathrm{Hz}$.  The central wells contained 
around 1500 atoms.

\begin{figure}[t]
\scalebox{0.47}{\includegraphics{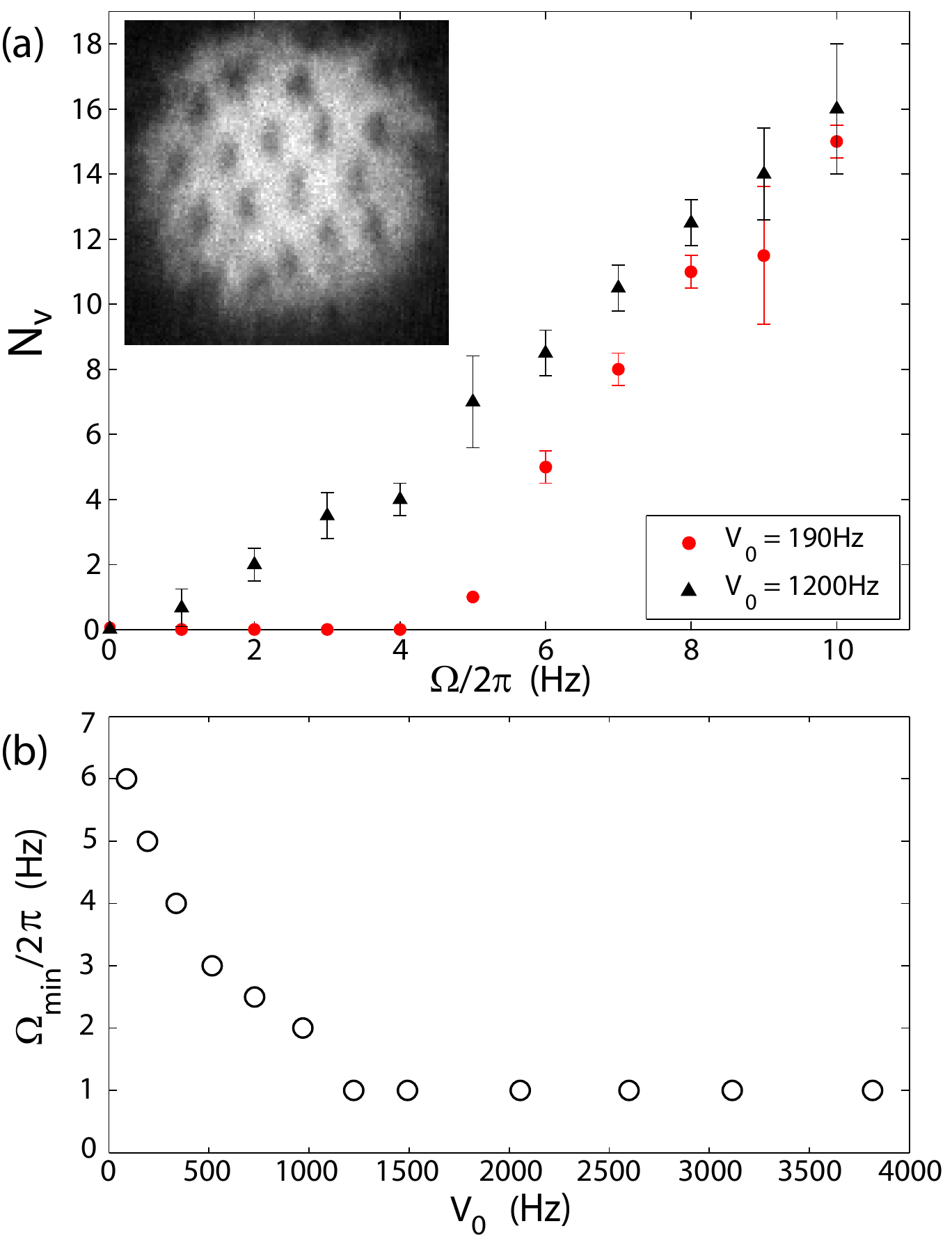}}
\caption{(a) Number of vortices nucleated as a function of optical lattice rotation frequency $\Omega$ for two lattice 
depths, $V_0 = 190~\mathrm{Hz}$ and $V_0 = 1200~\mathrm{Hz}$.  The error bars denote the standard deviation in vortex 
number for three experimental images.  Inset: an example of a vortex lattice created by the rotating optical lattice of 
depth $V_0 = 1200~\mathrm{Hz}$ at $\Omega\slash 2\pi = 10~\mathrm{Hz}$. (b) The minimum rotation frequency needed to 
nucleate a single vortex, $\Omega_{\mathrm{min}}$, as a function of the lattice depth $V_0$ (1~Hz was the minimum 
frequency used).}
\label{fig:omega_min_plus_Nv}
\end{figure}

After the condensate had been loaded into the static optical lattice $\Omega$ was increased linearly from zero to its 
final value in 320~ms, followed by a further 100~ms of rotation at the final value.  While the lattice was still 
rotating the lattice depth was then ramped down to zero in 12~ms, allowing the condensates to merge and converting 
phase differences around plaquettes to vortices.  Previously Scherer {\it{et al}} \cite{scherer07} have investigated 
vortex formation by merging three independent condensates, and Schweikhard {\it{et al}} \cite{schweikhard07} have 
observed the thermal activation of vortices on a two-dimensional lattice of Josephson-coupled BECs, however our 
experiment takes place in the rotating frame.  Rotation gives rise to an additional phase $2\pi f$ around a plaquette 
of the optical lattice, simulating the effect of a magnetic field on charged particles.  After the optical lattice was 
ramped down the cloud was then immediately released from the magnetic and optical traps and destructively imaged after 
20~ms time-of-flight expansion.  We also repeated the sequence but holding the cloud after the lattice was ramped down 
for an additional 500~ms, to allow any vortex-antivortex pairs to annihilate (such pairs could be created in a static 
lattice \cite{schweikhard07}).  This generally improved visibility of vortices in the harmonic trap and we observed the 
same number of vortices in the system within experimental error.  The uncertainty in the number of vortices observed 
for a given $\Omega$ arose from difficulties in determining the presence of vortices near the edge of a condensate.

\begin{figure}[t]
\scalebox{0.40}{\includegraphics{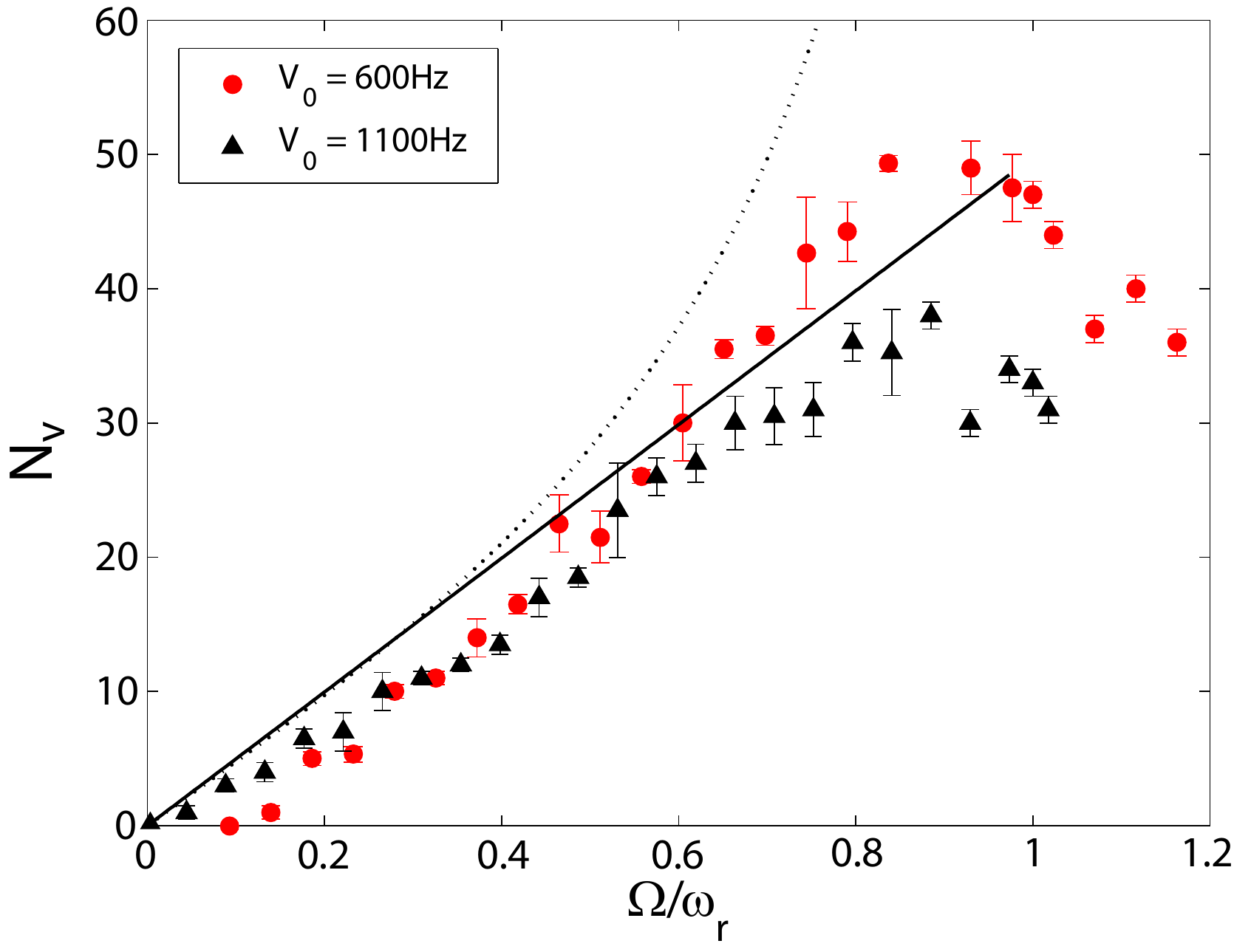}}
\caption{Number of vortices nucleated as a function of $\Omega\slash\omega_r$.  The solid line is the number of 
vortices expected for a cloud of fixed radius, $R_0 = 16~\mu\mathrm{m}$, while the dashed line in the number of 
vortices expected for a centrifugally distorted cloud of radius $R = R_0[1 - (\Omega\slash\omega_r)^2]^{-3\slash 10}$.  
We observed the optical lattice prevented the cloud from reaching $R$ for a given $\Omega$, allowing rotation above the 
critical frequency $\Omega > \omega_r$ to be achieved.}
\label{fig:Nv_vs_omega_0to20Hz}
\end{figure}

Figure \ref{fig:omega_min_plus_Nv}(a) shows the number of vortices, $N_v$, observed as a function of the optical 
lattice rotation frequency, $\Omega$, for lattice depths of $V_0 = 190~\mathrm{Hz}$ and $V_0 = 1200~\mathrm{Hz}$.  The 
behaviour of $N_v(\Omega)$ has a characteristic shape which depends on the exact vortex nucleation process.  The first 
BEC stirring experiments used stirring beams of a similar size to the condensate and observed that no vortices were 
nucleated below $\Omega \approx 0.7\omega_r$, where there is excitation of the $l = 2$ surface mode \cite{madison00}.  
It was later shown that sufficiently small stirring beams nucleate vortices without discrete resonances of surface 
modes \cite{raman01}, and the minimum frequency for vortex nucleation, $\Omega_{\mathrm{min}}$, matched the frequency 
at which a single vortex at the center of a condensate is energetically stable, $\Omega_c$ \cite{fetter_svid_review01}.  
For our experimental conditions $\Omega_c\slash{2\pi} \approx 4~\mathrm{Hz}$.  For $V_0 = 190~\mathrm{Hz}$, far below 
the $\approx 1000~\mathrm{Hz}$ needed to split the condensate into weakly linked islands, Fig.\ 
\ref{fig:omega_min_plus_Nv}(a) shows a similar trend to that observed in \cite{raman01}, i.e.\ $N_v$ did not exhibit 
sharp resonances and $\Omega_\mathrm{min} \approx \Omega_c$.  We checked this behaviour was an equilibrium property by 
repeating the experiment with rotation times up to 1~s, and observed the same response.  We infer that a weak ($V_0\ll 
\mu$) rotating lattice stirs up a condensate in a non-resonant fashion.  For a rotating lattice of depth $V_0 = 
1200~\mathrm{Hz}$, however, $N_v$ displayed a distinctly different trend, with the number of vortices observed 
following a linear dependence on $\Omega$ even below $\Omega_c$.  This change in behaviour is indicative of entering 
the regime where the system forms an array of Josephson-coupled condensates, rather than simply a condensate perturbed 
by a weak rotating lattice as for $V_0 = 190~\mathrm{Hz}$.  Figure \ref{fig:omega_min_plus_Nv}(b) highlights the 
transition between the two regimes, showing the dependence of $\Omega_{\mathrm{min}}$ on $V_0$.  
$\Omega_{\mathrm{min}}$ falls below $\Omega_c$ as $V_0$ rises above $\mu = 500~\mathrm{Hz}$, and falls to 1~Hz (the 
minimum rotation frequency used) for $V_0\geq 1000~\mathrm{Hz}$, corresponding to the lattice depth at which 
condensates are expected to be well localized on sites and can only communicate by tunneling.  The rotation induced 
phase differences between neighbouring condensates are converted to vortices when the lattice is ramped down in the 
rotating frame.  This nucleation mechanism creates vortices near the centre of the condensate even when $\Omega < 
\Omega_c$. Vortices do not have to spin in from the edge of the cloud, which is what we observed for stirring with a 
weak lattice potential.

The measurements of the number of vortices $N_v$ was extended to higher rotation frequencies $\Omega$ as shown in 
Figure \ref{fig:Nv_vs_omega_0to20Hz}. The ratio $\Omega\slash\omega_r$ is used where $\omega_r$ is the radial trapping 
frequency taking into account the slight enhancement by the Gaussian envelope of the lattice beams: for $V_0 = 
600~\mathrm{Hz}$, $\omega_r/2\pi = 21.5~\mathrm{Hz}$ and for $V_0 = 1100~\mathrm{Hz}$, $\omega_r/2\pi = 
22.6~\mathrm{Hz}$.  The density of vortices for a given $\Omega$, whether a lattice is present or not, is $n_v = 
m\Omega\slash\pi\hbar$.  The total number of vortices is then $N_v = m\Omega R^2\slash\hbar$.  For a rotating BEC in 
equilibrium at $\Omega$ the centrifugal distortion of the Thomas-Fermi radius is $R = R_0[1 - 
(\Omega\slash\omega_r)^2]^{-3\slash 10}$ \cite{fetter09}.
The predicted number of vortices for a cloud displaying this behaviour is denoted by the dashed line in 
Fig.~\ref{fig:Nv_vs_omega_0to20Hz}.  The solid line presumes the radius stays constant at $R_0=16~\mu\mathrm{m}$.  For 
$V_0 = 1100$~Hz, $N_v$ rises linearly with $\Omega$, whereas for the weaker $V_0 = 600$~Hz lattice $N_v$ shows a 
faster, nonlinear dependence on $\Omega$.  We attribute this behaviour to the deeper lattice suppressing the effect of 
centrifugal distortion.  For the deeper lattice atoms may only redistribute themselves by tunneling between sites.  
While the tunneling parameter (the Josephson-coupling energy \cite{kasamatsu09}) is around $2~\mathrm{kHz}$ at central 
lattice sites this drops to $<50~\mathrm{Hz}$ at peripheral lattice sites due to the drop in atom number, giving 
tunneling times at the edge of the lattice comparable to the experimental duration.  Bloch oscillations may also play a 
role in inhibiting the atoms from spreading out in the lattice \cite{wang06}.

\begin{figure}[t]
\scalebox{0.63}{\includegraphics{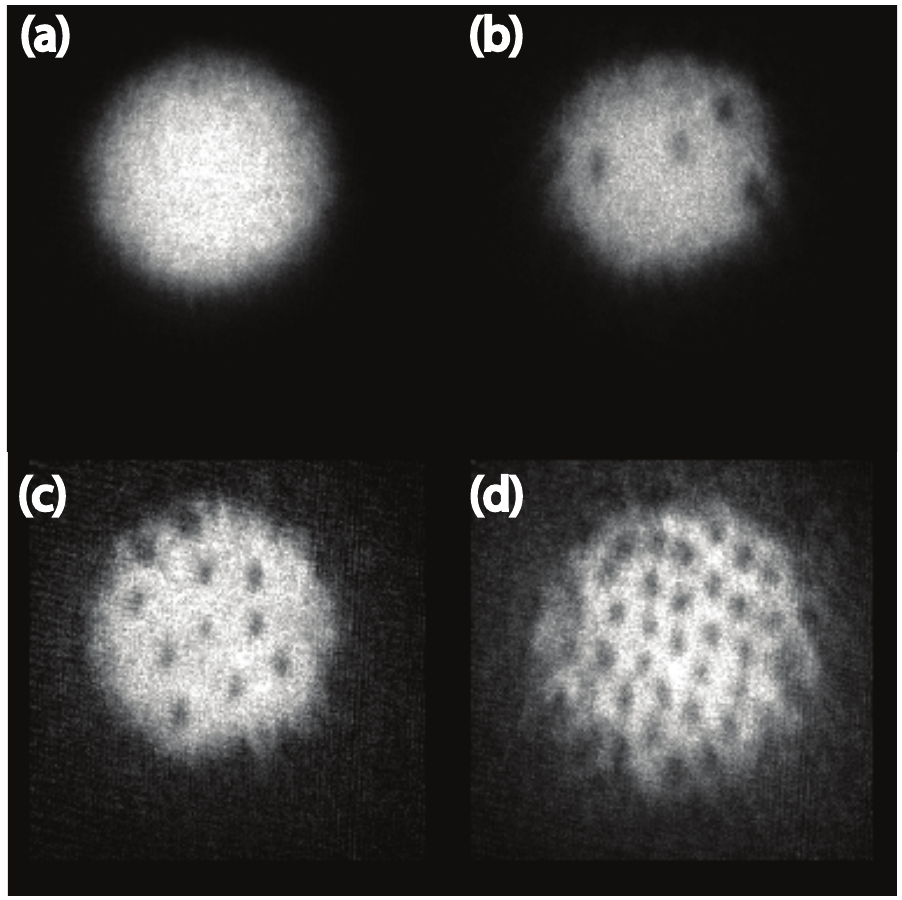}}
\caption{Images after release from the rotating optical lattice showing the pattern of vortices: (a) Condensate before 
rotation; (b) $\Omega\slash 2\pi = 4$~Hz; (c) $\Omega\slash 2\pi = 7$~Hz; (d) $\Omega\slash 2\pi = 17$~Hz.}
\label{fig:vortex_montage}
\end{figure}

As a result of the optical lattice inhibiting the spreading out of the cloud we were able to rotate the cloud above the 
critical frequency $\Omega > \omega_r$.  As seen in Fig.~\ref{fig:Nv_vs_omega_0to20Hz} the number of vortices dropped 
for rotation above $\omega_r$, due to increased heating of the condensate.  However vortices could be observed up to 
$\Omega = 1.15~\omega_r$ for $V_0 = 600~\mathrm{Hz}$ before heating and reduction of vortex visibility meant it was no 
longer possible to observe vortices.  The heating at $\Omega\sim\omega_r$ was worse for the deeper 
$V_0=1100~\mathrm{Hz}$ lattice and vortices could not be observed above $\Omega=1.02\omega_r$.  A cloud in a harmonic 
trap without the presence of an optical lattice should be expelled for $\Omega > \omega_r$ though Bretin {\it{et al.}}\ 
were able to achieve $\Omega = 1.06~\omega_r$ using additional quartic confinement \cite{bretin04}.

The structure of the vortex patterns formed in a bosonic JJA under rotation are heavily influenced by the presence of 
the lattice.  For values of the frustration parameter (the mean number of vortices per lattice plaquette) $f= 2m\Omega 
d^2\slash h=p\slash q$ (where p and q are integers) the ground state vortex patterns have been calculated 
\cite{kasamatsu09}, and have a unit cell of size $q\times q$ sites.  Between these rational fraction values of $f$ 
intermediate patterns are formed.  Figure \ref{fig:vortex_montage} shows examples of the vortex patterns we observe.  
We do not reproducibly observe the predicted $q\times q$ unit cell structure as we scan $\Omega$ and hence $f$.  There 
are several possible reasons for this: (i) for $d = 2~\mu\mathrm{m}$ and $\omega_r\slash 2\pi = 20.1~\mathrm{Hz}$ the 
maximum value of $f$ we can reach is $f\sim 1\slash 5$.  The finite size of the system then becomes an issue, as it 
will be difficult to realize the periodic structure for $q = 5$ when a maximum of 200 sites are filled by the 
condensate; (ii) the energy landscape of cold atoms in a rotating lattice is known to be very glassy, that is closely 
spaced minima in energy can be separated by large barriers, making it hard to reach the true ground state 
\cite{goldbaum09}.  Lack of equilibration on experimental timescales might then account for the absence of $q\times q$ 
vortex structures; (iii) any intrinsic heating from the rotating optical lattice will make it difficult to reach the 
ground state.

The issue of finite size can be addressed by increasing $\omega_z$ or using a larger condensate, enabling more lattice 
sites to be filled.  The values of $f$ achievable can be increased by implementing tighter radial trapping to allow 
higher $\Omega$ to be reached or increasing the lattice constant $d$.  The $f = 1\slash 2$ case is particularly 
interesting due to competition between an Ising-type and BKT type phase transition \cite{polini05}.  Vortex patterns in 
a bosonic JJA under rotation for $0 \leq f \leq 1\slash 2$ will be explored in future work.

\begin{figure}[t]
\scalebox{0.36}{\includegraphics{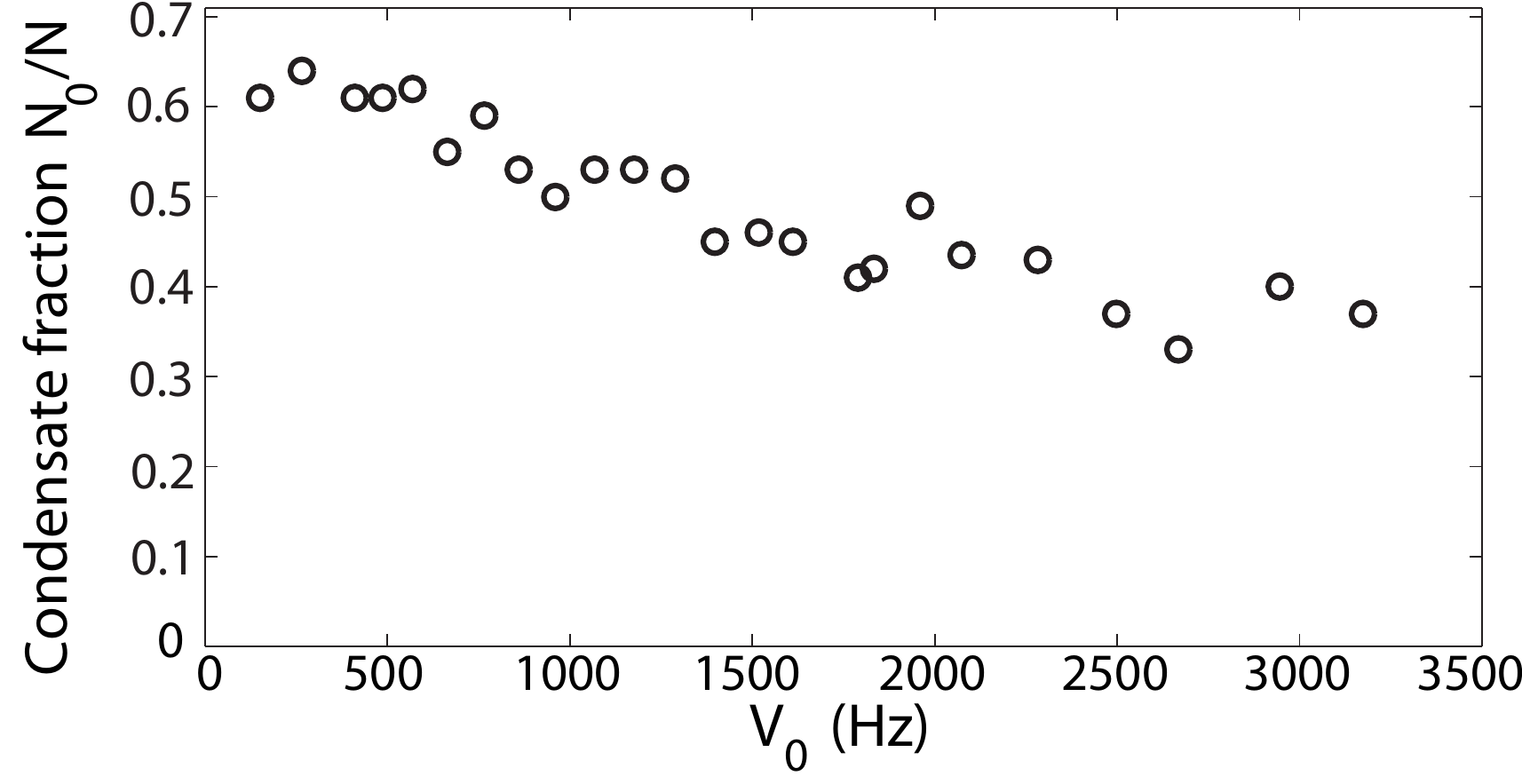}}
\caption{Condensate fraction $N_0\slash N$ as a function of lattice depth $V_0$ after ramping up the rotation rate to 
15~Hz in 40~ms followed by 35~ms of rotation at 15~Hz.}
\label{fig:heating}
\end{figure}

An important question regarding the viability of future rotating optical lattice experiments is the issue of heating.  
Figure~\ref{fig:heating} displays the fraction of atoms in the condensate, $N_0\slash N$, after rotation at 
$\Omega\slash 2\pi = 15~\mathrm{Hz}$ ($\Omega\slash 2\pi$ was increased from 0 to 15~Hz in 40~ms followed by 35~ms of 
rotation at 15~Hz) for different lattice depths, showing an approximately linear dependence.  We found at $\Omega\slash 
2\pi = 15~\mathrm{Hz}$, $V_0 = 1100~\mathrm{Hz}$ we could rotate for around 300~ms before heating significantly reduced 
the condensate fraction, with longer rotation times achievable for smaller $\Omega$ or $V_0$ or both.  The main sources 
of heating were fluctuations in the lattice depth during rotation (at the $\pm 1\,\%$ level) and residual imperfections 
in the lattice upon rotation.  All the data in this paper were taken without using any additional evaporation during or 
after the rotation process (unlike previous stirring experiments \cite{madison00, aboshaeer01}).  The implementation of 
an additional cooling scheme in the optical lattice such as that described by Griessner {\it{et al}}.\ 
\cite{griessner06} could enable sufficiently low temperatures to be achieved to realise quantum Hall physics in future 
experiments with the rotating lattice \cite{palmer:180407, sorenson05, bhat:043601}.

In conclusion, the use of acousto-optic deflection to generate smooth rotation of an optical lattice at a well-defined 
frequency has allowed us to explore a new method of vortex nucleation.  For lattice depths $V_0<\mu$ we found the weak 
lattice acted as a stirring mechanism.  For deeper lattices the system was split into well-localized condensates at 
each site and a linear dependence of vortex number on $\Omega$ was observed for rotation frequencies below the 
thermodynamic critical frequency for vortex nucleation. This implied vortices were created locally at a lattice 
plaquette.
\begin{acknowledgments}
This research is supported by EPSRC, QIP IRC (GR/S82176/01) and ESF.
\end{acknowledgments}


\begin{thebibliography}{30}
\bibitem[{not()}]{noteV0}
\bibinfo{note}{$V_0$ was calibrated by Kapitza-Dirac diffraction with
  an uncertainty of $\pm 10\%$. The measured $V_0$ agreed with that calculated
  from measured parameters within this error.}
\end{thebibliography}
\end{document}